# Thermal Characterization of Microscale Heat Convection under Rare Gas Condition by a Modified "Hot Wire" Method

Jianshu Gao[1,2], Yanan Yue[1,2*], Danmei Xie[1,2], Yangheng Xiong[1,2]

[1]State Laboratory of Hydraulic Machinery Transients (Wuhan University), MOE, Wuhan, 430072, China

[2]School of Power and Mechanical Engineering, Wuhan University, Wuhan, Hubei, 430072, China

**Abstract:**

As power electronics shrinks down to sub-micron scale, the thermal transport from a solid surface to environment becomes significant. Under circumstances when the device works in rare gas environment, the scale for thermal transport is comparable to the mean free path of molecules, and is difficult to characterize. In this work, we present an experimental study about thermal transport around a microwire in rare gas environment by using a steady state "hot wire" method. Unlike conventional hot wire technique of using transient heat transfer process, this method considers both the heat conduction along the wire and convection effect from wire surface to surroundings. Convection heat transfer coefficient from a platinum wire in diameter 25 μm to air is characterized under different heating power and air pressures to comprehend the effect of temperature and density of gas molecules. It is observed that convection heat transfer coefficient varies from 14 W/m$^2$K at 7 Pa to 629 W/m$^2$K at atmosphere pressure. In free molecule regime, Nusselt number has a linear relationship with inverse Knudsen number and the slope of 0.274 is employed to determined equivalent thermal dissipation boundary as $7.03 \times 10^{-4}$ m. In transition regime, the equivalent thermal dissipation boundary is obtained as $5.02 \times 10^{-4}$ m. Under a constant pressure, convection heat transfer coefficient decreases with increasing temperature, and this correlation is more sensitive to larger pressure. This work provides a pathway for studying both heat conduction and heat convection effect at micro/nanoscale under rare gas environment, the knowledge of which is essential for regulating heat dissipation in various industrial applications.

---

* Corresponding author: Yanan Yue, E-mail: yyue@whu.edu.cn

## 1. Introduction

As dimension of micro-electronic continues to scale down and the power density is enhanced, valid thermal management for effective heat dissipation becomes a crucial problem [1,2]. Various efforts have been devoted to enhancing heat conduction of solid or interface materials [3,4]. However, the impact of heat convection is of equal importance as scale shrinks to micro/nanoscale, where the characteristic length for thermal transport is comparable with mean free path of gas molecule [5]. It is evidenced that when the size of materials shrinks to microscale, convection heat transfer coefficient can be several orders of magnitude compared with that at macroscale [6]. For example, for materials with high specific surface area such as microwires, heat convection should not be ignored and it may account for a large proportion even in rare gas condition [7]. However, quantitative analysis and thermal characterization for understanding heat dissipation involving both heat conduction and convection effect at microscale, especially for extreme conditions such as rare gas atmosphere, is not sufficient.

Thermal transport in convection depends on not only heat conduction between hot surface and gas but also buoyance force driven by temperature difference and viscosity of the gas. However, buoyance force is negligible when the sample size shrinks to micro/nanoscale or is under rare gas atmosphere, thus heat transport is indeed dominated by heat conduction [8]. Heat loss through convection around material at micro/nanoscale is influenced strongly by the size of characteristic length and mean free path of gas molecule. Flow regime is classified into four types by Knudsen number ($Kn=\lambda/D$)[9]: continuum regime ($Kn<0.01$), slip regime ($0.01<Kn<0.1$), transition regime

($0.1<Kn<10$) and free molecule regime ($Kn>10$). In free molecule regime ($Kn>10$), the gas density is small so that gas molecules hardly collide with each other. Heat convection process is attributed to collisions between gas molecules and hot surface [10,11].

Thermal transport around wires with different diameters at various pressures was studied by Cheng et al.. The convection heat transfer coefficient is sensitive to pressure at pressure lower than 0.1 Torr. For pressure more than 100 Torr, heat convection is more closely related to sample diameter [12]. The work by Wang et al. reported a model for convection heat transfer coefficient prediction between carbon nanotube and gas environment which is valid for both free molecule regime and continuum regime [13]. Convection heat transfer coefficient around microwire with different diameters (10-100 μm) at atmosphere pressure was studied using 3ω method by Wang et al.. The surface to volume ratio is increased for microwires so that convection heat transfer coefficient at microscale is several orders of that at macroscale but the heat conduction along the wire is neglected [14]. Convection heat transfer coefficient was measured by using nanoscale heaters in different sizes. It is observed that convection heat transfer coefficient is sensitive to the heater size [15]. By using infrared thermometer, local temperature of micro-device was determined to characterize convection heat transfer coefficient. The low kinetic energy of air molecules promotes heat convection by diffusion process [16].

In this work, we present an experimental study about thermal transport around a microwire in rare gas environment by using a steady-state "hot wire" method. Convection heat transfer coefficient at different pressures nearly vacuum condition is characterized. Thermal diffusion boundary is determined from free molecule regime to transition regime to analyze how heat carriers diffuse and collide with each other, which is meaningful to the thermal design of micro/nanoelectronics.

## 2. Experimental principles and details

### 2.1 Experimental principles

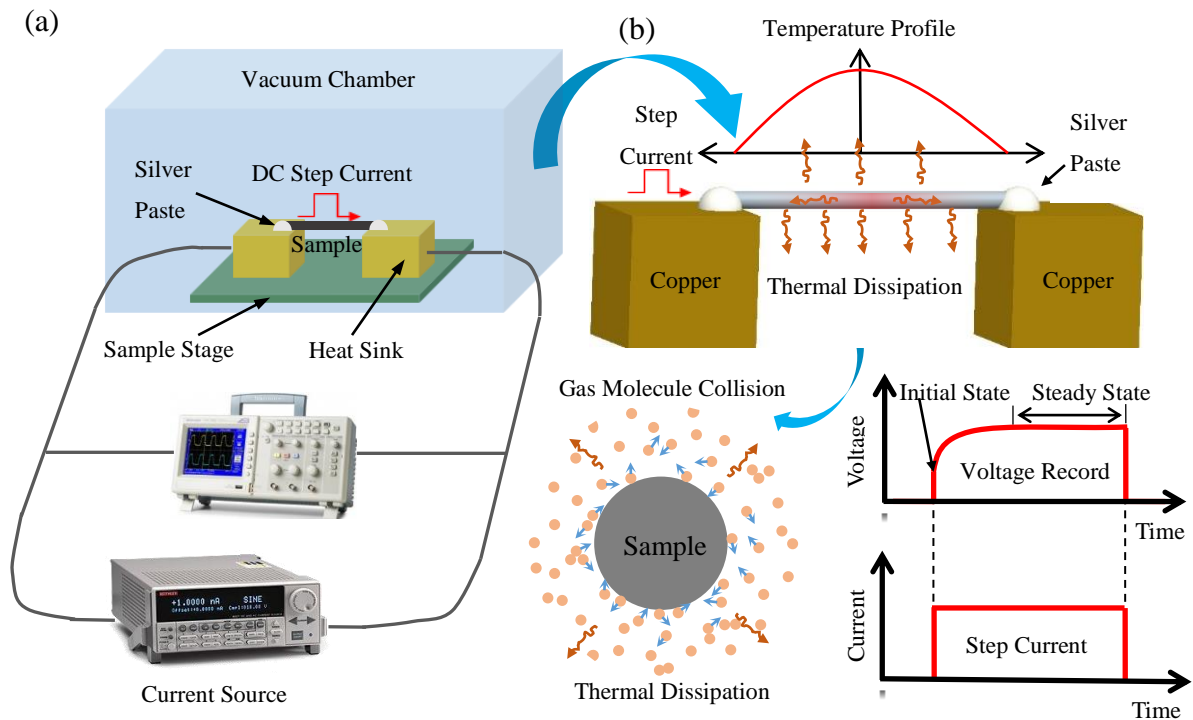

Figure 1. (a) Schematic setup for measuring convection heat transfer coefficient in rare gas environment. The sample is suspended between two heat sinks in a pressure adjustable chamber. A DC step current is introduced by current source and voltage variation is recorded for temperature measurement. (b) Thermal dissipation

inside the wire though heat conduction, and from wire surface to surrounding air by heat convection and heat radiation. The sample temperature reaches a steady state when thermal transport balances heat generation.

Fig. 1(a) illustrates the schematic setup for this steady-state "hot wire" method. The to-be-measured wire is suspended between two heat sinks in a pressure adjustable steel chamber. A DC step current is applied to wire to introduce electrical heating. The joule heat is generated inside the sample and dissipates through the wire and from wire surface to surrounding air as shown in Fig. 1(b). Heat conduction and convection as well as radiation effect are all involved to determine temperature profile along the wire. Temperature rise of the wire reaches a steady state when thermal dissipation balances heat generation inside the wire. We choose a wire material with its electrical resistance being very sensitive to temperature. The wire itself is employed as a temperature sensor so that the voltage variation recorded during heating process can provide temperature information. Once the temperature information is obtained, convection heat transfer coefficient is determined by solving one-dimensional conventional heat conduction equation as $\partial^2 T / \partial x^2 + q_0 / k = 0$. Here $T$ is temperature along the sample, $k$ is thermal conductivity, $x$ is the distance away from the sample middle point and $q_0$ is heat generation per unit volume from joule heating combined with the term of heat loss to surrounding air, and is described as $q_0 = Q/(A_c L) - 4h_e(T-T_0)/D$. Where $Q$ is joule heating power, $A_c$ is cross-section area of the sample, $L$ is the length, $T_0$ is room temperature and $h_e$ is effective convection heat transfer coefficient from the wire surface to surrounding air (combining with the influence of heat convection and heat radiation). The ends of the sample are assumed to be at room temperature, the temperature profile can be solved as

$$T(x) = \frac{Q}{h_e LS}(1 - \frac{e^{\sqrt{h_e Sx/kA_c}} + e^{-\sqrt{h_e Sx/kA_c}}}{e^{\sqrt{h_e SL/2kA_c}} + e^{-\sqrt{h_e SL/2kA_c}}}) + T_0 \quad (1)$$

where $S$ is wire perimeter. The average temperature rise of wire is determined as

$$\Delta \bar{T} = \frac{Q}{h_e LS} - \frac{2Q}{h_e L^2 S\sqrt{h_e S/kA_c}} \tanh(\frac{L\sqrt{h_e S/kA_c}}{2}) \quad (2)$$

when considering radiation heat loss, convection heat transfer coefficient is obtained as $h = h_e - \varepsilon\sigma(\bar{T}^4 - T_0^4)/(\bar{T} - T_0)$ Where $\varepsilon$ is reflection coefficient (0.05 for the material used in our experiment) [17], $\sigma$ is the Stefan-Boltzmann constant. Assuming that thermal conductivity remains constant for a small temperature range, convection heat transfer coefficient can be obtained from average temperature rise of wire.

This "hot wire" method is similar to conventional method in experimental setup, but different in measurement principle [18,19]. Conventional hot wire method is a transient process, and heat conduction along the wire is neglected. The slope for wire temperature is used for characterizing thermal property of materials around it. Since temperature profile along the wire is not uniform, heat conduction and convection process are all involved. In our steady-state "hot wire" method, heat conduction along the wire is counted when calculating convection heat transfer coefficient which makes the measurement more accurate.

The mean free path of gas molecule (λ) in this work is served as $\lambda = k_B T_{m,DF} / (\sqrt{2}\pi d^2 P)$ [20,21]. Where $k_B$ is the Boltzmann constant, $T_{m,DF}$ is the effective mean temperature of λ, $d$ is the molecular diameter, $P$ is the gas pressure in chamber.

## 2.2 Experimental details

The platinum wire used in this experiment has a diameter of 25 μm and length of 19.44 mm. The wire diameter is much smaller than the size of chamber. Thus the wire is supposed to be among infinite gas environment. Platinum wire possess very stable thermal conductivity (71.6 W/mK) within not very large temperature range [22]. With a high ratio of length over diameter, the wire can be well applied to one-dimensional model. The wire is suspended between two copper electrodes with ends attached to the electrodes by silver paste (to reduce contact resistance). Copper electrodes are much larger than sample dimension, and are used as heat sinks. Thus, the sample ends can be regarded being at room temperature (or a little temperature rise which is also considered in the data processing to reduce the uncertainty brought by such effect). The measurement is conducted in a vacuum chamber where pressure is adjustable from 0.1 Pa to atmosphere pressure. In the measurement, the wire is heated under different DC step currents (from 10 mA to 86 mA). Various Joule heating brought different temperature rise and it can be used to study the temperature effect on convection heat transfer coefficient. It is noted that small currents would bring large noise-to-signal ratio, which leads to large uncertainty in determining heat convection coefficient. To minimize such effect, the currents are supplied by a very high precise current source (KEITHLEY 6220) and all data curves are averaged more than 5 times for curve fitting by using Eq. (2).

## 3. Results and discussions

**3.1 Convection coefficient of microwire in a wide pressure range at the same temperature**

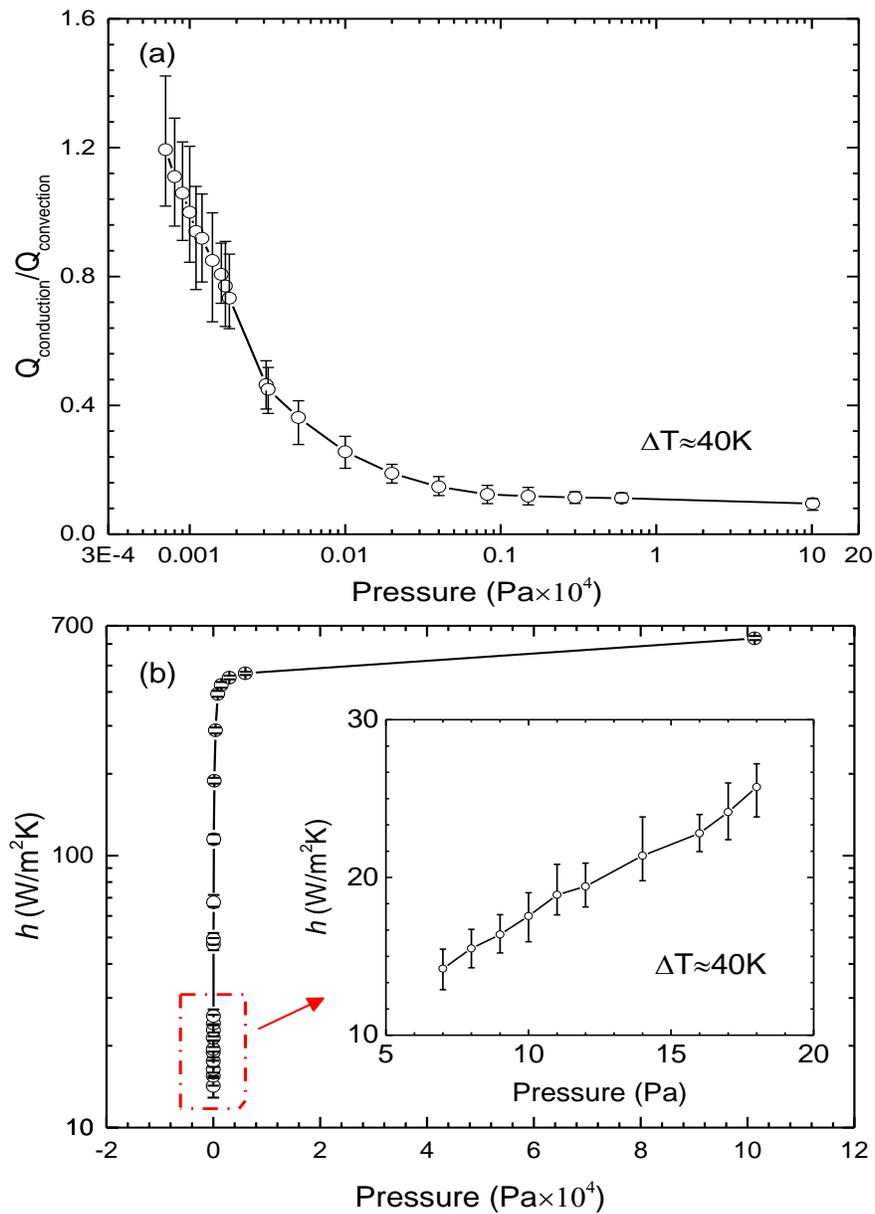

Figure 2. (a) The ratio of heat conduction to heat convection for a wire under different pressures. Heat conduction takes a large proportion, and is even more significant than convection in heat dissipation process at lower pressure. Since heat convection is improved as pressure is increased, heat conduction still holds a ratio of 9% even at atmosphere pressure. (b) $h$ under different pressures at a constant wire temperature. $h$ increases rapidly at low pressure, and then slowly toward atmosphere pressure. At low pressures, it shows a very linear correlation.

Within heat dissipation process along the wire, heat conduction accounts for a larger proportion than heat convection at pressure lower than 10 Pa (the ratio of heat conduction to heat convection is more than 1 in Fig. 2(a)). More heat is dissipated through convection so that the ratio gradually decreases as pressure is increased. However, heat conduction along the wire cannot be ignored since the ratio still holds 9% at atmosphere pressure. Pressure dependent convection heat transfer coefficient ($h$) at a constant temperature is shown in Fig. 2(b). Under rare gas condition, buoyance force is not significant for microwire [14], $h$ determined in this work is valid to different orientations. The proportion of radiation heat loss is less than 4% in effective convection heat transfer, which has been decoupled in calculating $h$. It shows that $h$ varies from 14 W/m$^2$K (7 Pa) to 629 W/m$^2$K (atmosphere), which is in good agreement with other references [14].

$h$ experiences a rapid increase at early pressure increment process, and then a slower increase toward atmosphere pressure. For pressure lower than around 28 Pa, thermal transport process is defined as ballistic transport that a gas molecule has little chance to collide with others in their migration after reflecting from the hot wire. $h$ increases rapidly with increasing pressure. When the pressure is more than 28 Pa, heat convection is dominated not only ballistic transport but also diffusion transport. The collisions between gas molecules increase and gradually play a significant role. At such, $h$ increases as pressure is increased, and this correlation is less sensitive to larger pressure when diffusion transport takes a larger proportion.

**3.2 Heat convection of microwire in free molecule regime under various heating**

Considering convection heat transfer in rare gas is different from that at atmosphere, the ballistic thermal transport in free molecule regime is introduced as [23]

$$Nu_{free} = (\frac{1}{\alpha_1} + (\frac{D_1}{D_2})(\frac{1}{\alpha_2} - 1))^{-1} \frac{(\gamma+1)}{Kn(9\gamma-5)} \sqrt{\frac{T_{m,DF}}{T_{m,FM}}} \quad (3)$$

Where $C_v$ is heat capacity of molecules, $\alpha_1$ is thermal accommodation coefficient (0.87 for the air molecule-platinum interaction) [24], $\alpha_2$ is thermal accommodation coefficient at thermal dissipation boundary, $D_1$ is wire diameter, $D_2$ is thermal dissipation boundary, $\gamma$ is specific heat ratio (1.4 for the air), $T_{m,DF}$ is effective mean temperature to calculate $Kn$ as mentioned before, and $T_{m,FM}$ is effective temperature for the gas. When temperature rise is small, $T_{m,DF}$ as well as $T_{m,FM}$ is equal to the average temperature of hot wire and surrounding air [21]. Convection heat transfer coefficient in free molecule regime can be derived as

$$h_{free} = (\frac{1}{\alpha_1} + (\frac{D_1}{D_2})(\frac{1}{\alpha_2} - 1))^{-1} \frac{k(\gamma+1)}{D_1 Kn(9\gamma-5)} \quad (4)$$

In Eq. (4), $D_2$ is highly related to $h$. This gives us an idea that by characterizing $h$, thermal boundary of heat convection can be determined.

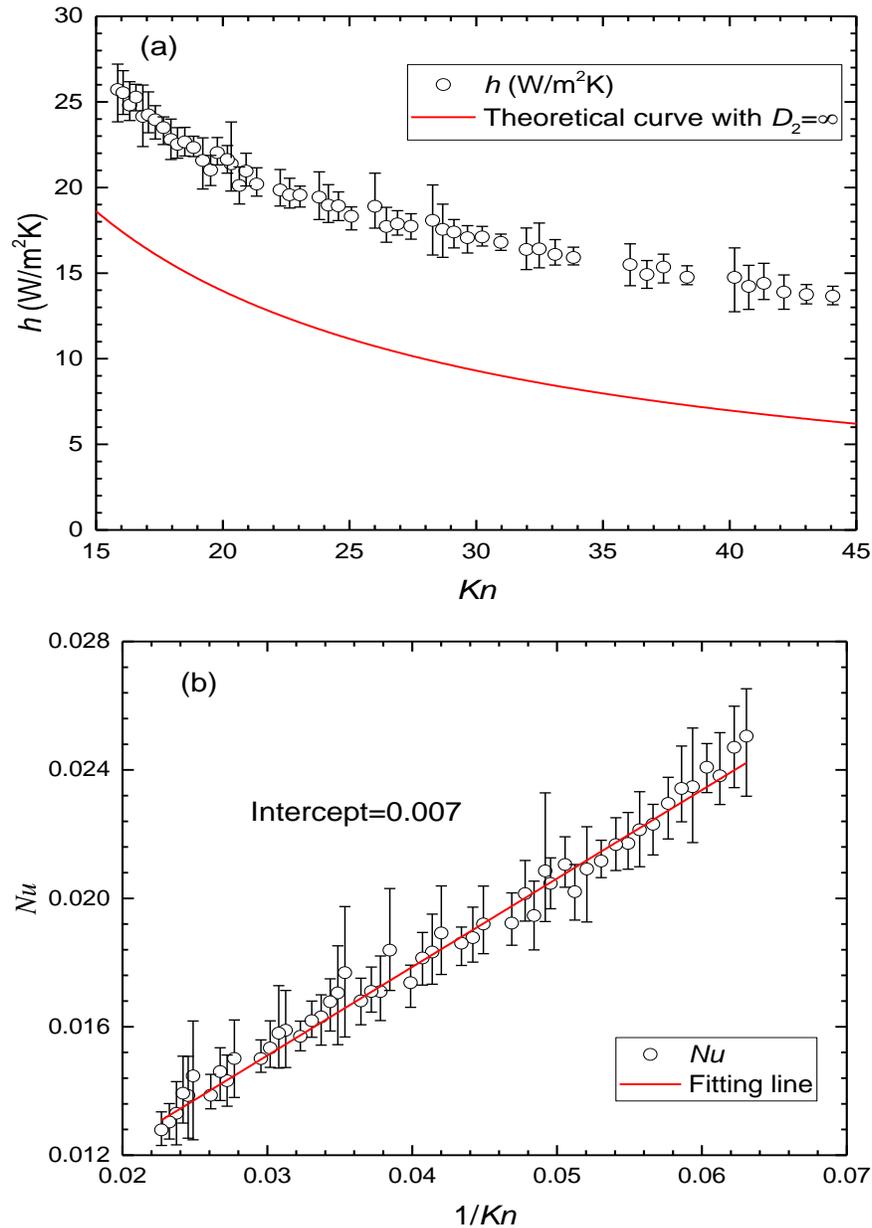

Figure 3. (a) $h$ under different $Kn$ in free molecule regime. A theoretical curve is introduced based on Eq. (4). The measured $h$ remains almost a constant larger than theoretically predicted values due to constant contact

thermal resistance. (b) *Nu* at various 1/*Kn* in free molecule regime. The intercept of Fitting line for *Nu* with respect to 1/*Kn* is determined as 0.007, which is useful for experimental data correction.

As demonstrated in Fig. 3(a), *h* decreases as *Kn* is increased in free molecule regime. It is reasonable that *Kn* is a reflection to the degree of rarefaction in the flow [25]. A larger value of *Kn* represents less gas molecules surrounding wire to take part in heat convection. A theoretical curve is introduced based on Eq. (4). It is found that *h* remains almost a constant larger than theoretically predicted values. Nusselt number (*Nu*) is sensitive to inverse Knudsen number (1/*Kn*) in free molecule regime in Fig. 3(b). The intercept of linear fitting for *Nu* with respect to 1/*Kn* is determined as 0.007. From Eq. (4) as derived, *Nu* is a linear relationship to 1/*Kn*, which means that the intercept of linear fitting for *Nu* with respect to 1/*Kn* should be zero. The non-zero error for the intercept in our measurement is from a constant contact resistance at the sample ends. Additional thermal resistance at contact point contributes to a small temperature rise of the wire, which leads to a larger value for calculating (apparent) convection heat transfer coefficients. Considering such effect, all values of *Nu* need to be corrected based on the intercept of fitting line in Fig. 3(b).

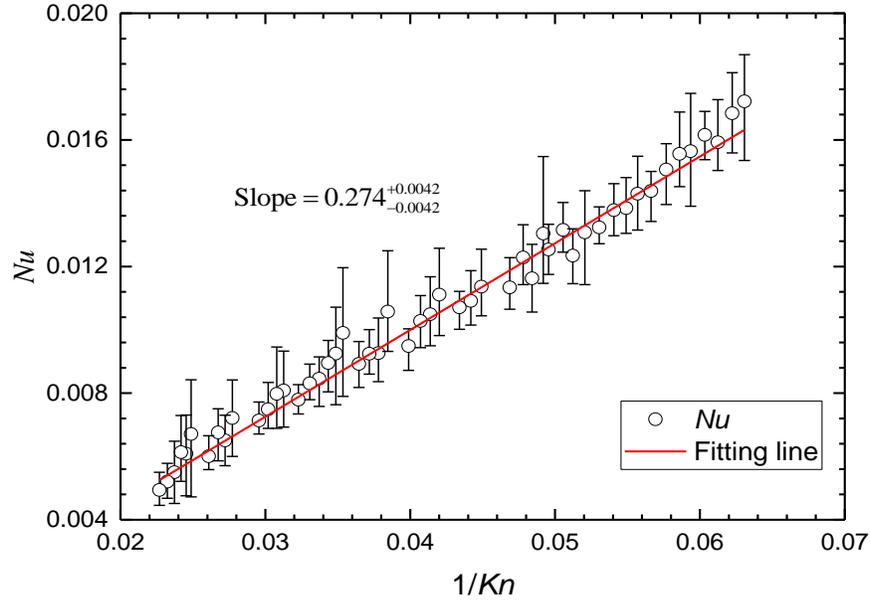

Figure 4. *Nu* after correction under diverse 1/*Kn* in free molecule regime. *Nu* shows a linear relationship to 1/*Kn* with a fitting slope of $0.274^{+0.0042}_{-0.0042}$. The equivalent thermal boundary is determined as $7.03 \times 10^{-4}$ m based on Eq. (4).

The *Nu* after correction is shown in Fig. 4. In free molecule regime, heat is carried by gas molecules away from hot wire through single collisions. Thermal transport is sensitive to 1/*Kn* with a slope fit as $0.274 \pm 0.0042$. Under circumstances when the thermal boundary is composed of air on steel with a thermal accommodation coefficient of 0.92 [21], the equivalent thermal boundary of transport process is determined as $7.03 \times 10^{-4}$ m based on Eq. (4). It means that the effect of heat dissipation in free molecule regime is from hot wire and ends at the distance of $7.03 \times 10^{-4}$ m. Note that the thermal boundary is closely related to wire diameter and the material at thermal boundary. The value of thermal boundary shrinks down several orders with a smaller wire diameter.

## 3.3 Heat convection of microwire in transition regime and its temperature effect

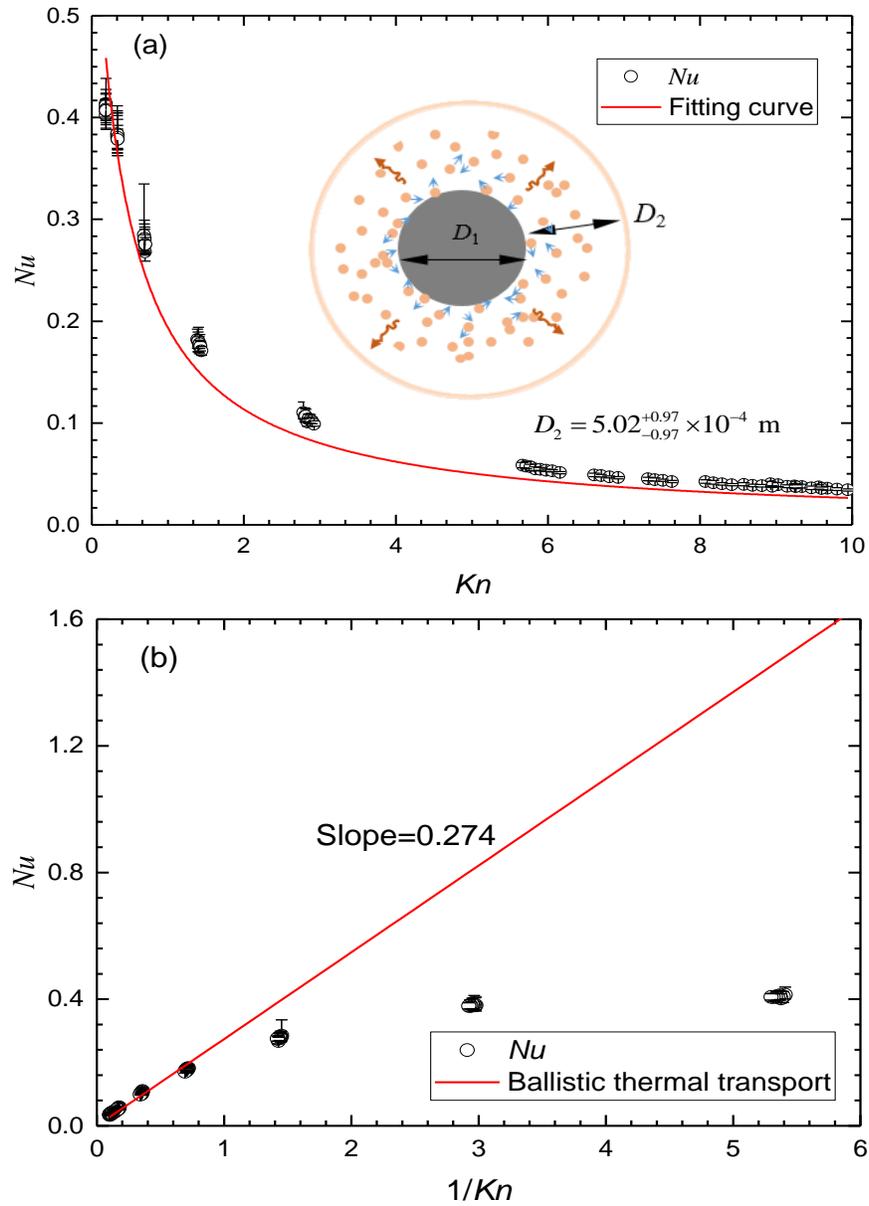

Figure 5. (a) $Nu$ under various $Kn$ in transition regime. The $Nu$ decreases as $Kn$ is increased. The equivalent thermal dissipation boundary is determined as $5.02_{-0.97}^{+0.97} \times 10^{-4}$ m, which is dozens of times compared with the wire diameter. (b) $Nu$ at various $1/Kn$ in transition regime. A line with a slope of 0.274 as that in free molecule

regime is introduced to represent ballistic thermal transport. The measured Nu in transition regime is smaller than ballistic thermal transport value.

For transition regime, heat dissipation process is occupied by ballistic thermal transport and diffusion thermal transport. Heat transfer through collision of gas molecules is described as [26,27]

$$Nu_{tran} = [1+\alpha_1 \frac{4B}{15} \frac{1}{2Kn} \ln(\frac{D_2}{D_1})]^{-1} Nu_{free} \qquad (5)$$

Where $B=1.184$ for a diatomic gas. $Nu$ decreases as $Kn$ is increased in transition regime, as shown in Fig. 5(a). A theoretical fitting curve is established to introduce thermal dissipation boundary based on Eq. (5). Thermal diffusion boundary is determined as $5.02\pm0.97\times10^{-4}$ m which is dozens of times compared with wire diameter. In Fig. 5(b), a line with a slope of 0.274 as that in free molecule regime is introduced to represent ballistic thermal transport. $Nu$ in transition regime is smaller than that predicted through ballistic thermal transport. In transition regime, diffusion thermal transport plays a significant role. Gas molecules which reflect from wire collide with other gas molecules, and then bounce back to wire surface again. The intermolecular collisions result in a lower $Nu$ in transition regime than that via ballistic thermal transport [28].

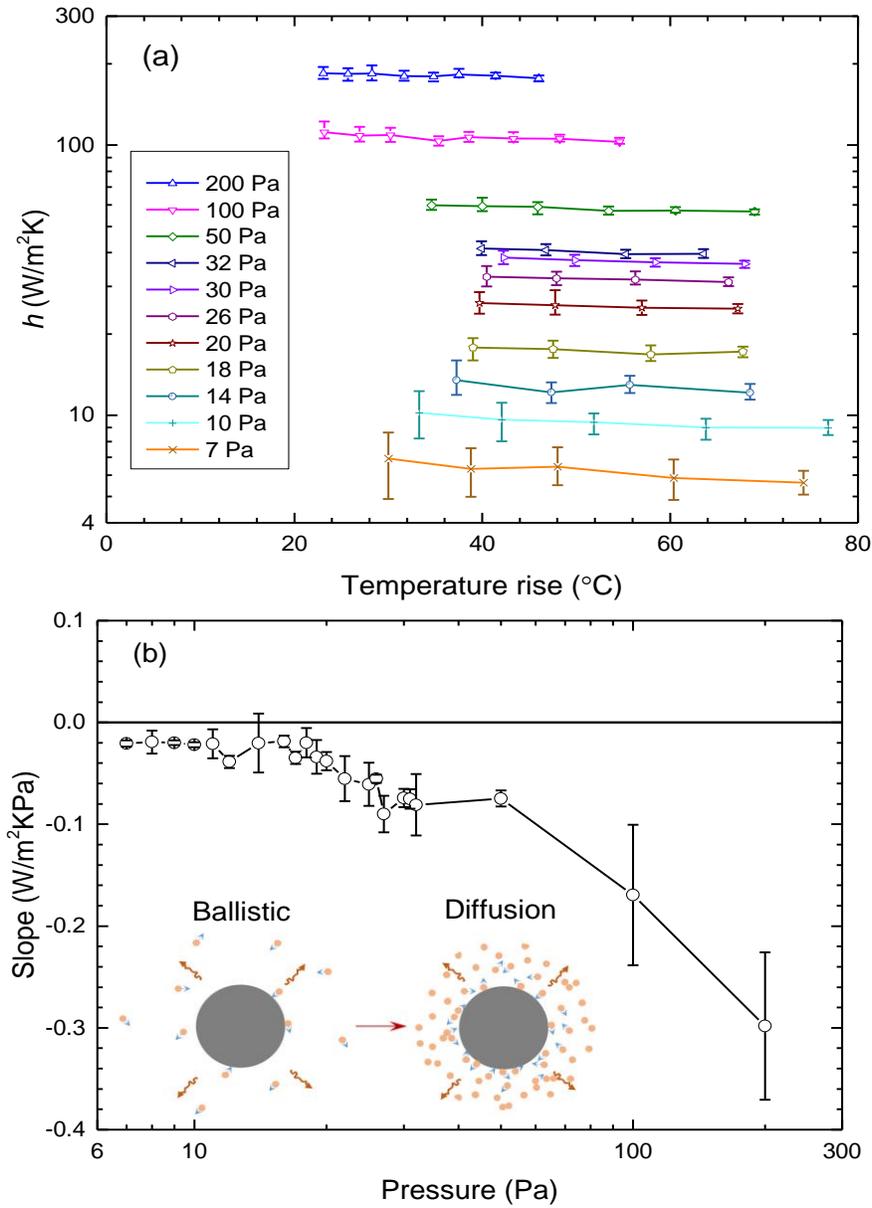

Figure 6. (a) The wire temperature effect on $h$ for different pressures. It shows that $h$ is decreased as temperature is increased at a certain pressure. (b) Slopes of $h$ with respect to temperature at a constant pressure. When pressure varies from 7 Pa to 200 Pa, heat transport evolves from ballistic thermal transport to diffusion thermal transport. The slope experiences a continuous decrease as pressure is increased.

Besides sample dimension and transport scales, temperature is another factor affecting heat convection effect significantly [29]. *h* is shown under different wire temperatures and gas pressures in Fig. 6(a). As pressure varies from 7 Pa to 200 Pa, *h* decreases as wire temperature is increased. The slopes of *h* with respect to temperature under different pressures in Fig. 6(b) experience a continuous decrease as pressure is increased. When pressure is increased, thermal transport evolves from ballistic thermal transport to diffusion thermal transport. Increased temperature enlarges *Kn* under a constant pressure, and thus weaken heat convection. This effect is enhanced as pressure is increased. Besides, as diffusion transport starts to be significant but not be the main portion, this correlation is still valid so that the slope continues to decrease.

### 3.4 Uncertainty analysis in measurement

Thermal conductivity of platinum wire is considered constant in this measurement. If there exist 5% difference (more or less) in thermal conductivity, the maximum error of *h* is 9% at lowest pressure or 0.26% at atmosphere pressure. It is mentioned that platinum wire possesses a very stable thermal conductivity within the temperature varying from 73 K to 1273 K [22]. Therefore, heat conduction along the wire can be counted precisely in this work. When using other wire materials either metals (such as copper microwire) or other lower-dimensional materials (such as carbon materials), different heating of materials certainly modifies thermal conductivity of inner structure and thus introduces large errors in determining *h* [29,22]. Great cautions are needed of this approach in such applications.

In addition, *h* obtained in this measurement is an averaged value along the wire. It is noticeable that the selection of a suitable wire length is important for characterizing local *h* on the wire. A long wire means heat convection takes a larger proportion in heat loss, and the effect of contact resistance may be neglected. However, it results in a larger difference to characterize local *h* with average *h* in our model since temperature difference along the wire is enlarged. This technique provides a pathway for studying both heat conduction and heat convection effect at micro/nanoscale in a very wide range from free molecule regime to continuum regime. The fluid around the wire can be changed into other liquids/rare gas. For materials with thermal conductivities unknown, we can obtain thermal property of solid wire and its temperature dependence in vacuum condition at first [30]. Then *h* under different heating power and pressures is obtained using our technique.

## 4. Conclusions

In summary, we characterize convection heat transfer coefficient of a microwire under different heating powers and pressures by using a modified (steady-state) "hot wire" method. Different from conventional hot wire technique, heat conduction along wire and convection effect are all involved to characterize thermal transport in this method, and more accurate and applicable in various industrial applications. Nusselt number is observed as linear relationship with inverse Knudsen number in free molecule regime. The equivalent thermal boundary is obtained in both free molecule and transition regimes. Under a certain pressure, convection heat transfer coefficient decreases with increasing temperature, and this correlation is more sensitive as pressure is increased. The newly developed method can be effectively used for studying micro/nanoscale heat convection effect at rare gas environment.

## Acknowledgements

The financial support from the National Natural Science Foundation of China (Nos. 51576145 and 51376140) are gratefully acknowledged. Authors also thank Mr Xiang Wan, Mr Changzheng Li and Miss Man Wang at Wuhan University for the help in the early experimental setup.